\documentclass[a4paper,11pt]{article}

\pdfoutput=1

\usepackage{jheppub}
\usepackage{amsfonts}
\usepackage{amsmath}
\usepackage{amssymb}
\usepackage{graphicx,color}
\usepackage{float}
\usepackage{hyperref}
\usepackage{subfigure}

\usepackage{xcolor}

\def\x{\mathbf{x}}

\setcitestyle{square, numbers, comma, sort&compress}

\title{\boldmath Disorder Effects in Dynamical Restoration of\\ Spontaneously Broken Continuous Symmetry
 }

\author[a]{Gustavo O. Heymans}    
\author[a]{, Nami F. Svaiter}    
\author[b]{and Gastão Krein}

\affiliation[a]{Centro Brasileiro de Pesquisas F\'{\i}sicas,
  Rua Xavier Sigaud 150, Rio de Janeiro, 22290-180 Rio de Janeiro, RJ, Brazil}
\affiliation[c]{Instituto de F\'{i}sica Te\'orica, Universidade Estadual Paulista,Rua Dr. Bento Teobaldo Ferraz, 271, Bloco II, 01140-070, S\~ao Paulo, SP, Brazil}

\emailAdd{olegario@cbpf.br}
\emailAdd{nfuxsvai@cbpf.br}
\emailAdd{gkrein@ift.unesp.br}

\abstract{We discuss the Euclidean quantum $O(N)$ 
model with $N=2$ 
in a continuous broken symmetry phase. We study the system at low temperatures in the presence of quenched disorder linearly coupled to the scalar field. Performing an average over the ensemble of all realizations of the disorder,
we represent the average free energy in terms of a series of the moments of the partition function. 
In the one-loop approximation, we prove that there is a denumerable collection of moments that lead the system to develop critical behavior. Our results indicate that in an equilibrium system, the strongly correlation of the disorder in imaginary produces generic scale invariance in the massive modes. } 
 

\begin{document}
\maketitle



\section{Introduction}\label{intro}

In the recent years, experimental advances in low temperature physics substantially increased the interest 
in the effects of noise and disorder in mesoscopic systems \cite {shakar,gustavo,gustavo3,zurek,sachdevbook,zvyagin, heyl}.
In this paper we study a continuous quantum field theory with quenched disorder linearly coupled to a scalar field. 
In a classical situation, a random field can model a binary fluid in porous media \cite{broch,dierker}.
When the binary-fluid correlation length is smaller than the porous radius, one has finite-size effects in 
the presence of a surface field. When the binary fluid correlation length is bigger than the porous radius, 
the random porous can exert a random field effect. In this case, the random field is linearly coupled with a 
classical field. The above example illustrates a general situation that  inhomogeneous backgrounds and
impurities can be modelled using random fields and random potentials within the formalism of continuous field theory. 
This issue raises fundamental questions regarding the role played by thermal, quantum and disorder induced fluctuations 
in a situation close to a second-order phase transitions.   

To drive the system to the criticality there are two quite distinct situations. The first one is when thermal or disorder-induced fluctuations are dominant. The second one is when quantum and disorder-induced fluctuations prevail over the thermal fluctuations \cite{hertz}. In systems at low temperatures, in the regime when the fluctuations' intrinsic frequencies $\omega$ satisfy $\omega\gg \beta^{-1}$, quantum dominate over thermal fluctuations.  In the case in which disorder-induced fluctuations prevail over thermal fluctuations, to investigate the low temperature behavior of the system one can work in the imaginary time formalism \cite{lebellac}. The steps are the following, first vacuum expectation values of operator products are continued analytically to imaginary time; then, under these analytically continued vacuum expectation values one imposes periodic boundary conditions in imaginary time. Finally, one can use functional methods where the finite temperature Schwinger functions are moments of a measure of some functional space \cite{yaglom1}.
 In such a situation of low temperatures, the disorder is strongly correlated in imaginary time. This situation has been studied in the literature, mainly in the context of the random mass model 
\cite{mvojta,vojta2,gr,gr2}. 

The algebraic decay of the correlation functions for generic control parameter values is known as generic scale invariance. For a spontaneously continuous broken symmetries, the presence of Goldstone modes is a signature of generic scale invariance, known as direct generic scale invariance. Nevertheless this is not the only way to produce generic scale invariance, for a review see Ref. \cite{belitz}.  In the case of discrete symmetry, the presence of quenched disorder also leads to the generic scale invariance. Such behavior is in agreement with Garrido \textit{et al} \cite{garrido}, who claim that a necessary, but not sufficient, condition for generic scale invariance is an anisotropic system. Some years latter Vespignani and Zapperi \cite{zapperi} showed that the breakdown of locality is essential to the generic scale invariance. In the the Ref. \cite{gustavo2} the authors proved that anisotropic disordr is a source of generic scale invariance.  The purpose of this paper is to generalize the results of Ref. \cite{gustavo2} for a  Euclidean quantum  
$O(N)$ model with $N=2$.
At low temperatures the system is in the ordered phase, with quantum and disorder induced fluctuations  prevailing over 
thermal fluctuations. In such a situation, we prove the appearance of indirect generic scale invariance.

To find the averaged free energy or generating functional of connected correlation functions, we use the distributional zeta-function method  \cite{distributional,distributional2,zarro1,zarro2,polymer,1haw,spin-glass}. After averaging the free energy over the disorder, we obtain a series representation in terms of the moments of the partition function.
Due to the strongly correlation of the disorder in imaginary time, there appears
a non-local contribution in each moment of the partition function. We circumvent the nagging nonlocality by using of the formalism of fractional derivative \cite{rr}.
We proved, in one loop approximation, that the system can make a transition from the ordered to the disordered phase by quantum and disorder induced fluctuations. We show that, below the critical temperature of the pure system, with the bulk in the ordered phase, there exist
a large number of critical temperatures that take each of these moments from an ordered to a disordered phase. This situation share some similarities with the Griffiths-McCoy phase, when in a quantum disordered system there appear finite size spatial regions in the disordered phase with bulk of the the system in the spontaneously broken, ordered, phase \cite{gri1,gri2}.

The structure of this paper follows. In Sec. \ref{sec:disoderedLG} we discuss the $O(N)$ for $N=2$ scalar field theory.
In Sec.  \ref{sec:frompathtoSPDE2} we study this model coupled
with quenched disorder. We use a series representation for the averaged free energy, the generating functional of connected correlation functions.
In Sec. \ref{sec:thermal mass2} we discuss
in the one-loop approximation
the effects of disorder in the broken symmetry phase.
We give our conclusions in Sec. \ref{sec:conclusions}. 
We use the units $\hbar=c=k_{B}=1$ throughout the paper.

\section{The Euclidean complex scalar Field}\label{sec:disoderedLG}

The action functional $S(\chi^{*},\chi)$ for an Euclidean complex scalar field  
at finite temperature in the imaginary time formalism \cite{cb,jackiw, rep} is given by  
\begin{align}\label{eq:1}
&S(\chi^{*},\chi)=\int_{0}^{\beta}d\tau\int d^{d}x
\left[\chi^{*}(\tau,\mathbf{x})\left(
-\frac{\partial^{2}}{\partial\tau^{2}}-\Delta+\mu_{0}^{2}\right)\chi(\tau,\mathbf{x})+\frac{\lambda }{4}| \chi^{*}(\mathbf{x})  \chi(\mathbf{x})|^{2} 
\right],
\end{align}
where $\beta$ is the reciprocal of the temperature, the symbol $\Delta$ denotes the Laplacian in 
$\mathbb{R}^{d}$, and $\lambda$ and $\mu_{0}^{2}$ are  respectively the bare coupling constant and the mass squared of the model. We omit subscripts indicating unrenormalized field and physical parameters, mass and coupling constant. The perturbative renormalization consists in the introduction of additive counterterms with coefficients $Z_{1}$, $\delta m^{2}$ and $Z_{2}$ to absorb the divergences in those quantities. 

The partition function is defined by the functional integral
\begin{equation}
Z=\int\left[ d\chi\right] \left[ d\chi^{*}\right]
\,\,\exp\bigl(-S(\chi^{*},\chi)\bigr),
\end{equation}
\\ 
where $\left[ d\chi\right] \left[ d\chi^{*}\right]$ is a functional measure and the field variables satisfy the periodicity condition $\chi(0,\mathbf{x})=\chi(\beta,\mathbf{x})$ and $\chi^{*}(0,\mathbf{x})=\chi^{*}(\beta,\mathbf{x})$. The generating functional of correlation functions is defined by introducing an external complex source $j(\tau,\mathbf{x})$ linearky coupled to the field.

We are interested in computing the ground state of the system in the situation where $O(2)$ symmetry is spontaneously broken. For accessing them, we replace $\mu^2_0$ by $-\mu^2_0$ in Eq. (\ref{eq:1}) and work with the Cartesian representation for the complex field $\chi(\tau,\mathbf{x})$. We define the real fields $\phi_{1}(\tau,\mathbf{x})$  and $\phi_{2}(\tau,\mathbf{x})$ such that
\begin{equation}\label{eq:ezk33}
\chi(\tau,\x)=\frac{1}{\sqrt{2}}\bigl[\phi_{1}(\tau,\x)+i\phi_{2}(\tau,\x)\bigr]
\end{equation}
and
\begin{equation}\label{eq:ezk333}
\chi^{*}(\tau,\x)=\frac{1}{\sqrt{2}}\bigl[\phi_{1}(\tau,\x)-i\phi_{2}(\tau,\x)\bigr].
\end{equation}
The potential contribution to the action functional, $V(\phi_{1},\phi_{2})$, is given in terms of these real fields by:
\begin{align}
\label{eq:effectivehamiltonianaaa}
V(\phi_{1},\phi_{2})=\left[-\frac{\mu^{2}_{0}}{2}\bigl(\phi_{1}^{2}+\phi_{2}^{2}\bigr)
+\frac{\lambda}{16}\bigl(\phi_{1}^{2}+\phi_{2}^{2}\bigr)^{2}\right].
\end{align}
The $O(2)$ symmetry corresponds to the invariance of the action under rotations in the real fields $(\phi_{1},\phi_{2})$ plane. The minima of this contribution  of $V(\phi_{1},\phi_{2})$, for $\mu_{0} > 0$ and  $\lambda>0$, are on the circle  with squared radius $v^{2}$, $\phi_{1}^{2}(\tau,\x)+\phi_{2}^{2}(\tau,\x) = v^{2}$, where $v^{2} =  \frac{4\mu^2_{0}}{\lambda}$. There is an infinite number of  degenerate ground states. This is the standard situation  of spontaneous symmetry breaking. Defining   $\varphi(\tau,\x) = \phi_{1}(\tau,\x)-v$ and $\psi(\tau,\x)=\phi_{2}(\tau,\x)$,  the action functional in terms of $\varphi$ and $\psi$, $S(\varphi,\psi)$ is given by 
\begin{align} \label{eq:effectivehamiltonian33}
S(\varphi,\psi) &= \int_{0}^{\beta}d\tau\int d^{d}x\,
\Biggl[\frac{1}{2}\varphi(\tau,\x)
\left(-\frac{\partial^{2}}{\partial\tau^{2}}-\Delta + m_{0}^{2}\right) \varphi(\tau,\x) +\lambda_{0}\Bigl(\varphi^{2} +\psi^{2}\Bigr)^{2} \nonumber \\ &+\frac{1}{2}\psi(\tau,\x) \left(-\frac{\partial^{2}}{\partial\tau^{2}}-\Delta\right) \psi(\tau,\x)
+\rho_{0}\varphi(\tau,\x)
\Bigl(\varphi^{2}+\psi^{2}\Bigr) \Biggr],
\end{align}
where we defined $m_{0}^{2}=2\mu_{0}^{2}$, $\lambda_{0} = \lambda/{16}$ 
and $\rho_{0} = \lambda v/{4}$. In the one-loop approximation it is possible to show that temperature restores the symmetry. In this case the ground state is unique.
The presence of the Goldstone-modes is a source of direct generic scale invariance. However we will go further and prove that the disorder field is able to generate indirect generic scale invariance. In the next section we introduce disorder in the system and discuss its effects in the restoration
of the spontaneously broken $O(2)$ symmetry.

\section{Euclidean complex scalar fields in Disordered Media}\label{sec:frompathtoSPDE2}

In this section we discuss the behavior of complex fields in disordered media. For the case of a statistical field theory with external randomness, one defines the action and has to find the quenched free energy, or the average of the generating functional of connected correlation functions in the presence of the disorder~\cite{englert, lebo,lebowitz}. 
 There are different ways to perform this average. Examples are the replica 
trick \cite{re,emery}, the dynamics approach \cite{dominicis, zip}, and the supersymmetry technique \cite{efe1}. Here we use the distributional zeta-function method \cite{distributional,distributional2,zarro1,zarro2,polymer,1haw,spin-glass}.

In a general situation, a disordered medium can be modelled by a real random field $\xi(\x)=\xi_{\omega}(\x)$
in 
$\mathbb{R}^{d}$  with $\mathbb{E}[\xi(\x)]=0$ and covariance $\mathbb{E}[\xi(\x)\xi(\mathbf{y})]$, where $\mathbb{E}[...]$ means average over an ensemble of realizations, i.e., over parameters $\omega$, characterising the disorder. In the case of a complex random field, the generalization is straightforward \cite{yaglom}. 
Let us consider a complex random field of real variables $h(x_{1},x_{2},...,x_{d})\equiv h(\x)$.
In the general situation we have:
\begin{equation}
\mathbb{E}[h(\x)]=m(\x),\,\,\,\,\,\mathbb{E}[h(\x)h^{*}(\mathbf{y})]=B(\x,\mathbf{y}),
\end{equation}
where $m(\x)$ and $B(\x,\mathbf{y})$ are respectively the first and the second moments of the random field. 
For simplicity we assume that the first moment is zero and the random field is delta correlated.
Therefore
the probability distribution of the disorder field is written as $[dh][dh^{*}]\,P(h,h^{*})$, where
\begin{equation}
P(h,h^{*})=p_{0}\,\exp\Biggl(-\frac{1}{2\,\varrho^{2}}\int\,d^{d}x|h(\x)|^{2}\Biggr),
\label{dis2}
\end{equation}
where $\varrho$ is a positive parameter associated with the disorder and $p_{0}$ is a normalization constant. In this case, we have a delta correlated disorder, i.e., $\mathbb{E}[{h^{*}(\x)h(\mathbf{y})}]=\varrho^{2}\delta^{d}(\x-\mathbf{y})$. The $[dh][dh^{*}]$ is a functional measure, where $[dh]=\prod_{\x} dh(\x)$.
The action functional in the presence of the complex disorder field is given by
\begin{align} 
&S(\chi,\chi^{*},h,h^{*})=S(\chi,\chi^{*}) + \, \int_{0}^{\beta} d\tau\int d^{d}x\,\Bigl(h(\x)\chi^{*}(\tau,\x)+h^{*}(\x)\chi(\tau,\x)\Bigr).
\end{align}
For examples of complex disordered fields, see Refs. \cite{tou,sham}.

One introduces the functional $Z(j,j^{*},h,h^{*})$, the disorder generating functional 
of correlation functions $i.e.$ generating functional of correlation functions for one 
disorder realization, where $j(\tau,\x)$ is an external complex source. 
As in the pure system case, one can define an average free energy as the 
average over the ensemble of all realizations of the disorder:
\begin{equation} 
\mathbb{E}\left[W(j,j^{*})\right] 
=\! \int\! [dh][dh^{*}]P(h,h^{*})\ln Z(j,j^{*},h,h^{*}).
\label{eq:disorderedfreeenergy}
\end{equation} 
%
%
%
%
The distributional zeta-function method computes this average of the free energy 
as follows. For a general disorder probability distribution, one defines the distributional 
zeta-function $\Phi(s)$:
\begin{equation}
\Phi(s)=\int [dh][dh^{*}]P(h,h^{*})\frac{1}{Z(j,j^{*},h,h^{*})^{s}}, \hspace{0.15cm}s\in \mathbb{C}, 
\label{pro1}
\vspace{.2cm}
\end{equation}
from which one obtains $\mathbb{E}\left[W(j,j^{*})\right]$ as
\begin{equation}
\mathbb{E}\bigl[W(j,j^{*})\bigr] = - (d/ds)\Phi(s)|_{s=0^{+}}, \,\,\,\,\,\,\,\,\,\, \Re(s) \geq 0.  
\end{equation}
%
Next, one uses Euler's integral representation for the gamma function $\Gamma(s)$ to write $Z^s$ as
\begin{equation} 
\frac{1}{Z^s} = \frac{1}{\Gamma(s)} \int^\infty_0 dt \, 
t^{s-1} \, e^{-Z \, t}
\end{equation}\\
Then, one breaks this $t$ integral into two integrals, one from $0$ to $a$ and another from 
$a$ to $\infty$, where $a$ is an arbitrary dimensionless real number, and expands the exponential 
into a power series of $t$ so that one can write the quenched free energy as
\begin{align}
\mathbb{E}\bigl[W(j,j^{*})\bigr] = \sum_{k=1}^{\infty} \frac{(-1)^{k+1}a^{k}}{k k!}\,
\mathbb{E}\,[Z^k(j,j^{*})] + \gamma   -  \ln(a)  + R(a,j,j^{*}),
\label{m23e}
\end{align}
\noindent 
where $\gamma$ is the Euler-Mascheroni constant~ \cite{abramowitz}, 
$\mathbb{E}\,[Z^{k}(j,j^{*})]$ is $k-$th moment of the partition partition 
function: 
\begin{align}
\hspace{-0.25cm}\mathbb{E}\,[Z^{\,k}(j,j^{*})] &= 
\int\,\prod_{i=1}^{k}[d\varphi_{i}^{(k)}]\prod_{j=1}^{k}[d\psi_{j}^{(k)}]
\exp\biggl(-S_{\textrm{eff}}(\varphi_{i}^{(k)},\psi_{j}^{(k)},j_{i}^{(k)},j_{j}^{(k)})\biggr),
\label{aa11}
\end{align}
in which the action $S_{\textrm{eff}}(\varphi_{i}^{(k)},\psi_{j}^{(k)})$ describes the field 
theory of $k-$field multiplets, and 
\begin{align}
R(a,j)=&-\int[dh] [dh^{*}]P(h,h^{*})
\int_{a}^{\infty}\,\dfrac{dt}{t}\, \exp\Bigl(-Z(j,j^{*},h,h^{*})t\Bigr).
\end{align}
For large $a$, $|R(a)|$ goes to zero exponentially, as shown in Ref. \cite{distributional}. 
Therefore, the dominant contribution to the average free energy is given by the moments of 
the partition function of the model. The effective action
$S_{\textrm{eff}}(\varphi_{i}^{(k)},\psi_{j}^{(k)},j_{i}^{(k)},j_{j}^{(k)})$ is a sum of
local and nonlocal terms with repect to the $\tau$ integral; we present them shortly 
ahead. The result of these steps is that, after the coarse-grained procedure with a 
reduced description of the disordered degrees of freedom, one gets collective variables 
that are multiplets of fields in the moments of the partition function. 

To proceed, we absorb $a$ in the functional measure and, following Klein and Brout, 
we define the augmented partition function $\mathcal{Z}(j,j^{*})$:
\begin{align}
\ln \mathcal{Z}(j,j^{*})=\sum_{k=1}^{\infty} c(k)\,\mathbb{E}\,[(Z(j,j^{*},h,h^{*}))^{\,k}].
\label{m23e}
\end{align}
where $c(k)=(-1)^{k+1}/{k k!}$.  Our purpose is to discuss the ordered phase of the model. 
Due a non-local contribution to the effective action, we restrict the discussion to equal fields in a given multiplet, that is: 
$\varphi^{(k)}_{i}(\x)=\varphi^{(k)}_{j}(\textbf{x})$, $\psi^{(k)}_{i}(\x) = \psi^{(k)}_{j}(\textbf{x})$ 
$\forall \,i,\,j$ in the function space. Likewise, we take $j_{i}^{(k)}(\textbf{x}) = j_{l}^{(k)}(\textbf{x})$. 
Therefore, all the terms of the series in Eq.~ \eqref{m23e} have the same structure. The effective action contain a local ($S_{\rm eff, L}$) and a non-local ($S_{\rm eff, NL}$) contribution, 
\begin{align}
&S_{\textrm{eff,L}}\left(\varphi_{i}^{(k)},j_{i}^{(k)}\right) = 
\frac{1}{2}\int_{0}^{\beta}d\tau\int d^{\,d}x \, 
\sum_{i=1}^{k} \Bigg\{ \varphi_{i}^{(k)}(\tau,\x) \left(-\frac{\partial^{2}}{\partial \tau^{2}}
- \Delta+m_{0}^{2}\right)\varphi_{i}^{(k)} \nonumber\\ &+ \, \psi_{i}^{(k)}(\tau,\x) \left(-\frac{\partial^{2}}{\partial \tau^{2}}
- \Delta\right)\psi_{i}^{(k)}(\tau,\x) 
+ \rho_0 \varphi_{i}^{(k)}(\tau,\x) \left[\bigl(\varphi_{i}^{(k)}(\tau,\x)\bigr)^2 
+ \bigl(\psi_{i}^{(k)}(\tau,\x)\bigr)^2 \right]
\nonumber \\ &+ \lambda_0\left[\left(\varphi_{i}^{(k)}(\tau,\x)\right)^2 
+ \bigl(\psi_{i}^{(k)}(\tau,\x)\bigr)^2 \right]^2
\Biggr\},
\label{SeffA}
\end{align}
\begin{align}
S_{\textrm{eff,NL}}\left(\varphi_{i}^{(k)},\psi_{i}^{(k)}\right) &=
-\frac{\varrho^{2}}{2\beta^2}\int_{0}^{\beta} d\tau\int_{0}^{\beta} d\tau'\int d^{d}x \, \sum_{i,j=1}^{k}
\left[\varphi_{i}^{(k)}(\tau,\x)
\varphi_{j}^{(k)}(\tau',\x) \right.\nonumber \\ &\left. +\, \psi_{i}^{(k)}(\tau,\x)
\psi_{j}^{(k)}(\tau',\x) \right].
\label{SeffA}
\end{align}
Here, we defined $\varphi'^{\,(k)}_{i}(\x)= 
 \frac{1}{\sqrt{k}}\varphi^{(k)}_{i}(\x)$, $\psi'^{\,(k)}_{i}(\x)= 
 \frac{1}{\sqrt{k}}\psi^{(k)}_{i}(\x)$, $\lambda'_{0}=\lambda_{0}k$ and $\rho'_{0}=\rho_{0}k$. As for the pure system, the fields variables satisfy periodicity condition in imaginary time: $\varphi_{i}^{(k)}(0,\x)=\varphi_{i}^{(k)}(\beta,\x)$ and $\psi_{i}^{(k)}(0,\x)=\psi_{i}^{(k)}(\beta,\x)$. In the context of replica trick, efforts was put to compute the contribution of the non-local action, in the replica symmetric case \cite{kirkpatrick} and in the replica symmetry break case, for a exemple see Refs. \cite{goldschmidt, azimi}. We would like to point out that, the restriction of equal fields in each multiplet is to avoid a technical problem raised by the non-local contribution. In the case of a local action one can perform the calculations choosing different fields in each multiplet in the functional space. 

However, in the context of distributional zeta-function method, the authors of Ref. \cite{gustavo2}, shows how the fractional derivative can deal naturally with the non-local contribution.  Through the Fourier transform, for a generic fractional derivative, we get
$\mathcal{F}\bigl[\frac{d^{\mu}g(x)}{d|x|^{\mu}}\bigr]=-|k|^{\mu}g(k)$, 
for $1\leq\mu<2$. The non-local contribution will appear in the Fourier representation of the Matsubara modes as $\frac{2\pi|n|}{\beta}$.



\section{Disorder and quantum effects in the one-loop contribution to the renormalized mass}\label{sec:thermal mass2}

The purpose of this section is to discuss the renormalized squared mass at very low temperatures. 
Since we use a regularization procedure where the Matsubara modes appear, we call it the thermal mass, 
although we are working at very low temperatures, in the spontaneously broken symmetry phase. We 
compute in the one-loop approximation the thermal mass in $k^{th}$ moment of the partition function. 
A Goldstone mass can not be generated in perturbation theory. Therefore, we concentrate on the
mass of the non-Goldstone mode.  There are two kinds of loops which give the first nontrivial 
contributions at the one-loop level; those with one insertion will be refereed to as tadpoles 
and with two insertions as self-energies. Below the critical temperature, one can write the 
square of the renormalized mass associated with the $k-$th momentum field $\varphi_{i}^{(k)}(\tau,\x)$
as:
\begin{align}
m_{R}^{2}(\beta,\varrho,k) &= m_{0}^{2} + \delta m^{2}_{0}(k) 
+ \, \lambda_{0} \Bigl[ 12I_{1}(\beta,\varrho,k)  + \, 2I_{1}(\beta,\varrho,k)|_{m^{2}_{0}=0} \Bigr] 
\nonumber \\ &- \, \rho_{0}^{2} \Bigl[ 18I_{1}(\beta,\varrho,k) +18I_{2}(\beta,\varrho,k)
+ \, 2I_{2}(\beta,\varrho,k)|_{m^{2}_{0}=0} + 6I_{1}(\beta,\varrho,k)|_{m_{0}^{2}=0}
\Bigr] 
\end{align}
where $\delta m_{0}^{2}(k)$ is the usual mass countertem that must be introduced in the renormalization procedure, for $k^{th}-$moment fields,  where the multiplicative numbers are symmetry factors, $I_1$ gives the tadpole contribution
\begin{align}
&I_{1}(\beta,\varrho,k)=\frac{1}{2L(2\pi)^{d}}
\int\,\prod_{i=1}^{d}dq_{i}\sum_{n\in \mathbb{Z}}\Biggl(q_{1}^{2}+...+q_{d}^{2}
+\biggr(\frac{2\pi n}{\beta}\biggl)^{2}+\frac{2\pi|n|}{\beta}\,k\varrho^{2}+m_{0}^{2}
\Biggr)^{-1}, 
\end{align}
and $I_2$ is the self-energy contribution
\begin{align}
I_{2}(\beta,\varrho,k)=\frac{1}{2L(2\pi)^{d}}
\int\,\prod_{i=1}^{d}dq_{i}\sum_{n\in \mathbb{Z}}\Biggl(q_{1}^{2}+...+q_{d}^{2}
+\biggr(\frac{2\pi n}{\beta}\biggl)^{2}+\frac{2\pi|n|}{\beta}\,k\varrho^{2}+m_{0}^{2}
\Biggr)^{-2}.
\end{align}
Note that we have infrared divergences for Nambu-Goldstone virtual loops. 
There are different ways to deal with those divergences~\cite{s1,s2,s3}. Here, we employ 
analytic regularization procedure~\cite{dim1,dim2,physica,bbb,ca12,ca13,ca14,ca15}. First, 
we define $\lambda(\mu,s)=\lambda_{0}(\mu^{2})^{s-1}$. Then, we define 
$\rho_{1}(\mu,s)=\rho_{0}(\mu^{2})^{s-1}$ and $\rho_{2}(\mu,s)=\rho_{0}(\mu^{2})^{s-2}$, 
where $\mu$ has mass dimension. Let us discuss first the $I_{1}(\beta,\varrho,k)$ integral.
We perform the angular part of the integral over the continuous momenta.  $I_{1}(\beta,\varrho,k)$ 
can be written as the analytical continuation of $I(\beta,\varrho,k,s)$ with $s\in \mathbb{C}$:
\begin{align}
I(\beta,\varrho,k,s)=\frac{\beta}{2^{d+2}\pi^{\frac{d}{2}+1}\Gamma\bigl(\frac{d}{2}\bigr)}
\int_{0}^{\infty}dp\,p^{d-1} \, \sum_{n\in \mathbb{Z}}\Biggl(\pi n^{2}+\frac{\beta}{2}k\varrho^{2}|n|+\frac{\beta^{2}}{4\pi}\Bigl(p^{2}+m_{0}^{2}\Bigr)
\Biggr)^{-s},
\end{align}
which converges for $\Re(s)>s_{0}$. Specifically, $I_1(\beta,\varrho,k)$ is given by the analytic continuation 
$I(\beta,\varrho,k,s)|_{s=1}$. Similarly, the finite part of the self-energy contribution is given by 
the analytical continuation $I_{2}(\beta,\varrho,k)=I(\beta,\varrho,k,s)|_{s=2}$.
To proceed, we define the dimensionless quantity 
$r^{2}={\beta^{2}p^{2}}/{4\pi}$. After a Mellin transform, and performing the~$r$~integral,  
$I(\beta,\varrho,k,s)$ is given by:
\begin{align}
&I(\beta,\varrho,k,s)=\frac{1}{8\pi\Gamma(s)}\biggl(\frac{1}{\beta}\biggr)^{d-1}\int_{0}^{\infty}dt\,t^{s-\frac{d}{2}-1}\sum_{n\in \mathbb{Z}}\exp{\Biggl[-\biggl(\pi\,n^{2}+
\frac{\beta}{2}k\varrho^{2}|n|+
\frac{m_{0}^{2}\beta^{2}}{4\pi}\biggr)t\Biggr]}.
\end{align}
Note that we are assuming at this point that $m_{0}^{2}\neq 0$. Let us split the summation into~$n=0$ and~ $n\neq 0$ contributions: 
\begin{equation}
I(\beta,\varrho,k,s)=
I(\beta,\varrho,k,s)|_{n=0}+I(\beta,\varrho,k,s)|_{n\neq 0}.
\end{equation}
The $n=0$ contribution is given by:
\begin{align}
I(\beta,\rho,k,s)|_{n=0}&=\frac{1}{8\pi\Gamma(s)}\biggl(\frac{1}{\beta}\biggr)^{d-1}A(s,d),
\end{align}
where $A(s,d)$ is 
\begin{align}
A(s,d)=\int_{0}^{\infty}dt\,t^{s-\frac{d}{2}-1}
\exp{\Biggl[-t\biggl(
\frac{m_{0}^{2}\beta^{2}}{4\pi}}
\biggr)\biggr].
\end{align}
The integral $A(s,d)$ is defined for $Re(s)>\frac{d}{2}$, and can be analytically 
continued to $Re(s)>\frac{d}{2}-1$ for $s\neq \frac{d}{2}$. Using the identity
\begin{align}
\int_{0}^{\infty}dt\,t^{s-\frac{d}{2}-1}
\exp{\Biggl(-t\biggl(
\frac{m_{0}^{2}\beta^{2}}{4\pi}}
\biggr)\Biggr)&=\int_{0}^{1}dt\, t^{s-\frac{d}{2}-1}
\Biggl[\exp{\biggl(-t\biggl(
\frac{m_{0}^{2}\beta^{2}}{4\pi}}
\biggr)\biggr)-1\Biggr]\nonumber \\ &+\int_{1}^{\infty}dt\, t^{s-\frac{d}{2}-1}
\exp{\Biggl(-t\biggl(
\frac{m_{0}^{2}\beta^{2}}{4\pi}}
\biggr)\Biggr)+\frac{1}{\bigl(s-\frac{d}{2}\bigr)},
\end{align}
which is valid for $Re(s)>\frac{d}{2}$. For $Re(s)>\frac{d}{2}-1$ and $s\neq \frac{d}{2}$, 
the right-hand side exists and defines a regularization of the original integral, that we
denote $A_{R}(s,d)$. The contribution $I(\beta,\varrho,k,s)|_{n\neq 0}$ is written as
\begin{align}
&I(\beta,\varrho,k,s)|_{n\neq 0}=\frac{1}{4\pi\Gamma(s)}\biggl(\frac{1}{\beta}\biggr)^{d-1}
\, \int_{0}^{\infty}dt\,t^{s-\frac{d}{2}-1}\, \sum_{n=1}^{\infty}\exp{\Biggl[-\pi t\biggl(n^{2}+
\frac{k \beta\varrho^{2}}{2\pi}n+\frac{m_{0}^{2}\beta^{2}}{4\pi^{2}}
\biggr)\Biggr]}.
\end{align}
Next, we make use of the properties of the Hurwitz-zeta function $\zeta(z,a)$
to handle this integral. $\zeta(z,a)$ is defined as 
\begin{equation}
\zeta(z,a)=\sum_{n=0}^{\infty}\frac{1}{\bigl(n+a\bigr)^{z}}, \,\,\,\,\,\,\,a\neq 0,-1,-2,...\, .
\end{equation}
for $z \in \mathbb{C}$. The series converges absolutely for ${\rm Re}(z)>1$. It is possible to find 
the analytic continuation with a simple pole at $z=1$. Inspired in the Hurwitz-zeta functions 
one defines the generalized Hurwitz zeta-function $Z(z,a)$ such that
\begin{equation}
Z(z,a)= \sum_{n=1}^{\infty}\frac{1}{\Bigl(\omega_{n}^{(k)}+a\Bigr)^{z}},
\end{equation}
for $a\notin (-\infty, -\omega_{n}^{(k)}]$ and $z \in \mathbb{C}$. 
Since in our case $a=\frac{m_{0}^{2}\beta^{2}}{4\pi}$ and $\omega_{n}^{(k)}=\pi n^{2}
+\frac{1}{2}\beta\varrho^{2} n k$, then one can write:
\begin{align}
I(\beta,\varrho,&k,s)|_{n\neq 0}=\frac{1}{4\pi}\biggl(\frac{1}{\beta}\biggr)^{d-1}\, 
\Gamma\biggl(s-\frac{d}{2}\biggr)\Bigr(\Gamma(s)\Bigl)^{-1}
\sum_{n=1}^{\infty}\frac{1}{\Bigl(\omega_{n}^{(k)}+a\Bigr)^{s-\frac{d}{2}}}.
\end{align}

We discuss further $I(\beta,\varrho,k,s)|_{n\neq 0}$ observing that a more general 
proof using generalized Hurwitz-zeta functions is based in the fact that zeta function 
regularization with a meromorphic extension to the whole complex plane needs an eligible 
sequence of numbers \cite{voros}. Therefore, in the series representation for the free 
energy with $k=1,2,..$, we have that for the moments of the partition function such that $k_{(q)}\leq \lfloor(\frac{2\pi q}{\beta})\frac{2}{\varrho^{2}}\rfloor$, 
where $(\frac{2\pi q}{\beta})$ are the positive Matsubara frequencies $\omega_{q}$, 
the system is critical for $q=\mathbb{N}$. This is an interesting result, since there 
are critical moments in the series representation for the free energy, after averaging 
the quenched disorder. Substituting the above result for $k_{(q)}= \lfloor(\frac{2\pi q}{\beta})\frac{2}{\varrho^{2}}\rfloor$, one gets that $I(\beta,q,s)|_{n\neq 0}$ 
can be written as
\begin{align}
I(\beta,q,s)|_{n\neq 0 } &= \frac{1}{4\pi\Gamma(s)}\biggl(\frac{1}{\beta}\biggr)^{d-1}
\int_{0}^{\infty}dt\,t^{s-\frac{d}{2}-1}\exp\left[
-\pi t\left(\frac{m_{0}^{2}\beta^{2}}{4\pi^{2}}-q^{2}\right)
\right] \nonumber\\
&\times\sum_{n=1}^{\infty}\exp{\Bigl[-\pi t\bigl(n+q
\bigr)^{2}\Bigr]}.
\end{align}
Finally, a simple calculation shows that choosing $q$ such that $q_{0}=\lfloor\frac{m_{0}\beta}{2\pi}\rfloor$, the quantity $
I(\beta,q,s)|_{n\neq 0}$ is given by:
\begin{align}
&I(L,q_{0},s)|_{n\neq 0} = \frac{1}{4\pi\Gamma(s)}\biggl(\frac{1}{\beta}\biggr)^{d-1}
\int_{0}^{\infty}dt\,t^{s-\frac{d}{2}-1}
\sum_{n=1}^{\infty}\exp{\Bigl[-\pi t\bigl(n+q_{0}
\bigr)^{2}\Bigr]}.
\end{align}
This simplification allows one to write $I(\beta,q_{0},s)|_{n\neq 0}$ as 
\begin{align}
&I(\beta,q_{0},s)|_{n\neq 0}=\frac{1}{4\pi\Gamma(s)}\biggl(\frac{1}{\beta}\biggr)^{d-1}
\Biggl[\int_{0}^{\infty}dt\,t^{s-\frac{d}{2}-1}
\sum_{n=0}^{\infty}\exp{\Bigl[-\pi t\bigl(n+q_{0}
\bigr)^{2}\Bigr]}-A_{R}(s,d)\Biggr].
\end{align}
Using the Hurwitz-zeta function and the integral $A_{R}(s,d)$
we can write
\begin{align}
&I(\beta,q_{0},s)|_{n\neq 0}=\frac{1}{4\Gamma(s)}\biggl(\frac{1}{\beta}\biggr)^{d-1}\Biggl[\frac{1}{\pi^{s-\frac{d}{2}+1}}
\Gamma\biggl(s-\frac{d}{2}\biggr)\zeta(2s-d,q_{0})-\frac{1}{\pi}A_{R}(s,d)\Biggr].
\end{align}

The contribution coming from the loops with Nambu-Goldstone bosons can be calculated, 
assuming $m^{2}_{0}=0$. All of these contributions must be regularized in the lower 
limit of the integrals. The Mellin transform of the contribution from the 
Nambu-Goldstone bosons is
\begin{align}
G(\beta,\varrho,k,s)|_{n\neq 0}&=\frac{1}{4\pi\Gamma(s)}\biggl(\frac{1}{\beta}\biggr)^{d-1}\lim_{m_0 \to 0}
\int_{0}^{\infty}dt\,t^{s-\frac{d}{2}-1}\nonumber \\ &\times \sum_{n=1}^{\infty}\exp{\Biggl[-\pi t\biggl(n^{2}+
k_{(q_0)} \frac{\beta\varrho^{2}}{2\pi}n + m_0^2\biggr)\Biggr]}.
\end{align}
Using the same regularization procedure we used to control infrared divergences, we 
obtain $[G(\beta,\varrho,k,s)|_{n\neq 0}]_{R}$. One can show that in low temperatures 
the contribution coming from $[G(\beta,\varrho,k,s)|_{n\neq 0}]_{R}$ is negligible. 
Now we will prove that for a fixed $q_{0}$, the renormalized squared mass vanishes 
for a family of $\beta's$. There are many critical temperatures where the 
renormalized squared mass is zero. We get that: 
\begin{align}
m_{R}^{2}(\beta,q_{0})&=m_{0}^{2}+\delta m_{0}^{2}+c_{1}\biggl(I(\beta,s=1)|_{n=0}\biggr)+c_{1}
\biggl(I(\beta,q_{0},s=1)|_{n\neq 0}\biggr)
\nonumber\\
&+c_{2}\biggl(I(\beta,s=2)|_{n=0}\biggr)+c_{2}
\biggl(I(\beta,q_{0},s=2)|_{n\neq 0}\biggr),
\end{align}
where $c_{1}=12\lambda(\mu,s)-18\rho_{1}^2(\mu,s)$  and 
$c_{2}=18\rho^{2}_{2}(\mu,s)$.
Defining the dimensionless quantity $b = m_0 \beta$, we write
\begin{align}
    \frac{b^{d-1}}{m_0^{d-3}} &- \frac{c_1}{8\pi}A_{R}(1,d) + \frac{c_{2}}{8\pi}A_{R}(2,d)  +\frac{c_1}{4\pi^{2-\frac{d}{2}}}\Gamma\left(1-\frac{d}{2}\right)\zeta\left(2-d,\frac{b}{2\pi}\right) \nonumber \\ &-\frac{c_{2}}{4\pi^{3-\frac{d}{2}}}\Gamma\left(2-\frac{d}{2}\right)\zeta\left(4-d,\frac{b}{2\pi}\right)+\delta m_{0}^{2}=0.
\end{align}
Let us discuss the important case where $d=3$. We get
\begin{align}
    &b^{2} - \frac{c_1}{8\pi}A_{R}(1,3) + \frac{c_{2}}{8\pi\mu^2}A_{R}(2,3) -\frac{c_1}{2}
    \zeta\left(-1,\frac{b}{2\pi}\right)
     -\frac{c_{2}}{4\pi}\lim_{d \to 3}\zeta\left(4-d,\frac{b}{2\pi}\right)+\delta m_{0}^{2} =0.
\end{align}
A formula that is relevant in the renormalization procedure is
\begin{equation}
\lim_{z\rightarrow 1}\biggl[\zeta(z,a)-\frac{1}{z-1}\biggr]=-\psi(a),
\label{psi}
\end{equation}
where $\psi(a)$ is the digamma function defined as
$\psi(z)=\frac{d}{dz}\bigl[\ln(z)\bigr]$.
The contribution coming from $A_R(s,d)$ is irrelevant for large $m_{0}\beta$. 
Using the identity $(n+1)\zeta(-n,a) = -B_{n+1}(a)$, where the $B_{n+1}(a)$ are the 
Bernoulli polynomials, we rewrite the Hurwitz-zeta function as
\begin{align}
    \zeta\left(-1,\frac{b}{2\pi}\right) &=
    -\left(\frac{b^2}{8\pi^2} - \frac{b}{4\pi} + \frac{1}{12}\right).
\end{align}
Using Eq. (\ref{psi}), we fix the counterterm contribution in the renormalization 
procedure. Then we have:
\begin{align}
    &b^{2} + \frac{c_1}{2}\left(\frac{b^2}{8\pi^2} - \frac{b}{4\pi} + \frac{1}{12}\right) +\frac{c_{2}}{4\pi}\psi\left(\frac{b}{2\pi}\right)\,=0.
\end{align}
Recognizing that $q_0=\lfloor\frac{b}{2\pi}\rfloor$, we can write
the digamma function as
\begin{align}
    \psi(q_0 + \alpha) = \psi(\alpha) + \sum_{q=1}^{q_0}\frac{1}{\alpha + q}\,,
\end{align}
where $\alpha$ is the non-integer part of $\frac{b}{2\pi}$. With $\alpha <1$ we can use a Taylor's series and write
\begin{align}\label{eq:b}
    & b^{2} + \frac{c_1}{2}\left(\frac{b^2}{8\pi^2} - \frac{b}{4\pi} + \frac{1}{12}\right) + \, \frac{c_{2}}{4\pi}\left(-\frac{1}{\alpha} - \gamma + \frac{\pi^2}{6}\alpha + H_{q_0}^{(1)} + \alpha H_{q_0}^{(2)}\right)=0,
\end{align}
where $H_{q_0}^{(1)}$ and $H_{q_0}^{(2)}$ are the generalized harmonic numbers. The above equation has zeroes for different values of $\beta$.
\begin{figure}[ht!]\label{fig:b}
    \centering\includegraphics[scale=1]{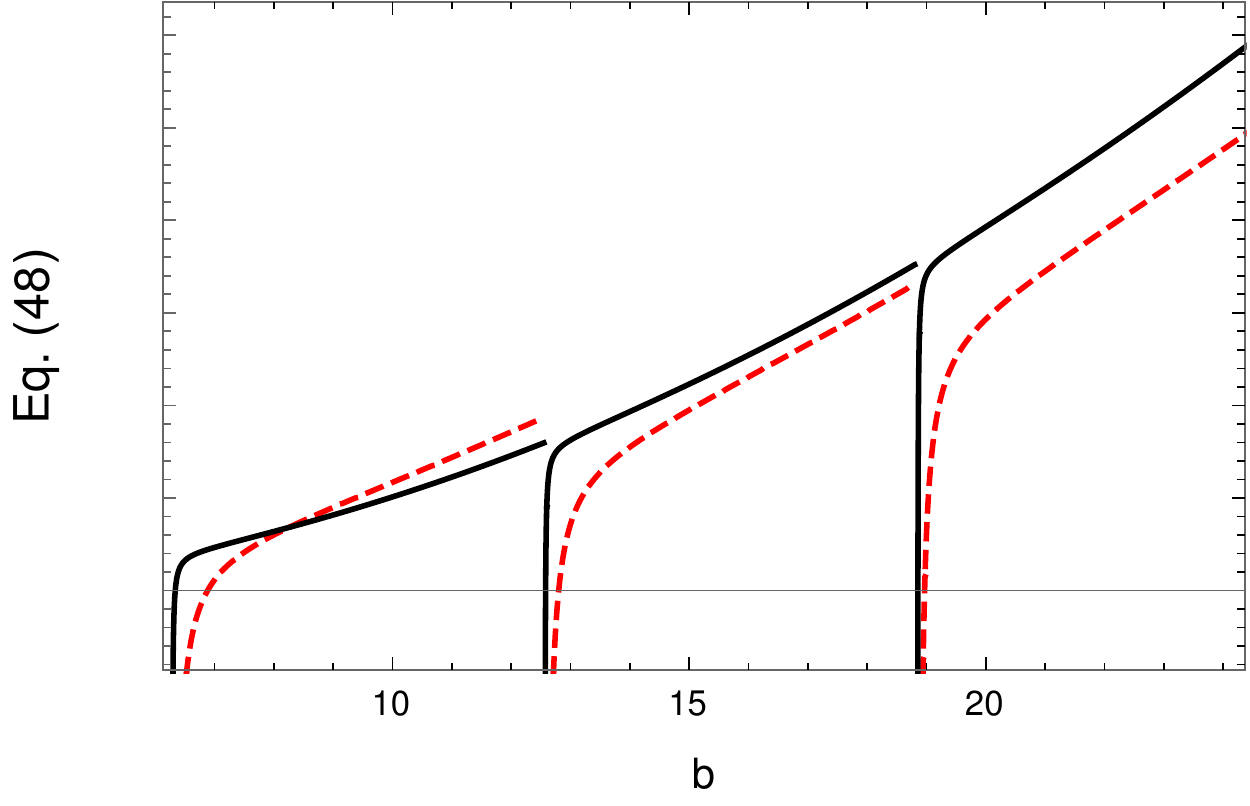}
 \label{fig:2}
\caption{Plot of Eq. (\ref{eq:b}) as a function of $b = m_0\beta$ for two different values of $\lambda$ (once $\rho_0^2 = \mu_0^2\frac{\lambda}{4}$): $\lambda = 1$ (continuous black) and $\lambda = 15$ (dashed red). We set $\mu^2=m_0^2$.}
\end{figure}

Our equation (47) and the Fig. 1, are manifestations of indirect generic scale invariance. In summary, we have proved in one-loop approximation that, in the set of moments that 
defines the quenched free energy, there is a denumerable collection of moments that 
can develop local critical behavior. Even in the situation where the bulk is in the ordered phase, temperature effects lead those moments from the ordered to a local disordered 
phase. This is our main result in this paper.

\section{Conclusions}
\label{sec:conclusions}

Recent experimental and theoretical advances increased activities in low temperature 
physics and quantum phase transitions. The intersection of these two area of research, 
the physics of quenched disordered systems and low temperatures lead to the following questions: how is the effect of randomness in the restoration of a spontaneously  broken continuous symmetry at low temperature? For a spontaneously continuous broken symmetries the presence of Goldstone modes is a signature of direct generic scale invariance. Here we discussed the consequences of introduce a random field in a system described by a complex field prepared in the ordered phase at low temperatures, with can be interpreted as the emergence of generic scale invariance. For a discrete symmetry the same behavior was obtained \cite{gustavo2}. 
In one-loop approximation they proved that with
the bulk in the ordered phase,
there is a denumerable set of moments that lead the system to this critical regime. 
In these moments appear a large number of critical temperatures.

We study the case where phase transitions is governed mainly by quantum and disorder induced fluctuations. The limit situation, when thermal fluctuations are absent is the case of 
quantum phase transitions. In this case the ground states of systems change in some fundamental way tuned by non-thermal control parameters of the systems.
Note that the low temperature behavior of the system, 
the disorder is strongly correlated in imaginary time.
Using the distributional zeta-function method after averaging the disorder under a coarse-graining, a non-local contribution appears in each effective action.  In one-loop approximation we discuss the effects of the disorder fluctuations in the restoration of the continuous broken symmetry by quantum and temperature effects. There are two results in the work. The first result is that we show that the contribution coming from the Nambu-Godstone loops is irrelevant to drive the phase transition. This is an expected result. In the one-loop approximation the criticality is obtained with the contribution coming from the thermal mass, and the Goldstone thermal mass can not be generated in perturbation theory.  The second one, in one-loop approximation, we proved that with
the bulk in the ordered phase,
there is a denumerable set of moments that lead the system to this critical regime. In these moments appear a large number of critical temperatures. This is a indication of indirect generic scale invariance in the system.

A natural continuation is show the result holds for high-order loops or even a non-perturbative regime, using a composite operator formalism \cite{cjt,pettini,gino}.

\begin{acknowledgments} 
 This work was partially supported by Conselho Nacional de Desenvolvimento Cient\'{\i}fico e Tecnol\'{o}gico - CNPq, 305894/2009-9 (G.K.), 303436/2015-8 (N.F.S.), INCT F\'{\i}sica Nuclear e Apli\-ca\-\c{c}\~oes, 464898/2014-5  (G.K) and Funda\c{c}\~{a}o de Amparo \`{a} Pesquisa do Estado de S\~{a}o Paulo - FAPESP, 2013/01907-0 (G.K).
G.O.H thanks Coordena\c{c}\~ao de Aperfei\c{c}oamento de Pessoal de Nivel Superior - CAPES 
for a PhD scholarship.  
\end{acknowledgments}
\appendix

\providecommand{\noopsort}[1]{}\providecommand{\singleletter}[1]{#1}%
\providecommand{\href}[2]{#2}\begingroup\raggedright\endgroup

\end{document}